\documentclass[fleqn,10pt]{wlscirep}
\usepackage[utf8]{inputenc}
\usepackage[T1]{fontenc}
\usepackage{newunicodechar}
\usepackage{bm}
\usepackage{graphicx}
\graphicspath{{images/}}
\usepackage{hyperref}
\title{A multitask framework for automated interpretation of multi-frame right upper quadrant ultrasound in clinical decision support}

\author[1*]{Haiman Guo}
\author[1*]{Cheng-Yi Li}
\author[1*]{Yuli Wang}
\author[2*]{Robin Wang}
\author[3,4]{Yuwei Dai}
\author[5]{Qinghai Peng}
\author[5]{Danming Cao}
\author[6]{Zhusi Zhong}
\author[7]{Thao Vu}
\author[3]{Linmei Zhao}
\author[8]{Chengzhang Zhu}
\author[1]{Christopher Tan}
\author[1]{Jacob Schick}
\author[1]{Stephen Kwak}
\author[1]{Farzad Sedaghat}
\author[1]{Javad Azadi}
\author[1]{James Facciola}
\author[1]{Jonathan Feng}
\author[1]{Dilek Oncel}
\author[1]{Ulrike Hamper}
\author[1]{Alex Zhu}
\author[1]{Tej Mehta}
\author[1]{Melissa Leimkuehler}
\author[1]{Cheng Ting Lin}
\author[6]{Zhicheng Jiao}
\author[3]{Ihab Kamel}
\author[9]{Jing Wu}
\author[4]{Li Yang}
\author[3**]{Harrison Bai}

\affil[1]{Department of Radiology, Johns Hopkins University School of Medicine, Baltimore, MD, USA}
\affil[2]{Department of Radiology, Stanford University School of Medicine, Stanford, CA, USA}
\affil[3]{Department of Radiology, University of Colorado Anschutz Medical Campus, Aurora, CO, USA}
\affil[4]{Department of Neurology, Second Xiangya Hospital, Central South University,Changsha, Hunan, China}
\affil[5]{Department of Computer Science, Brown University, Providence, RI, USA}
\affil[6]{Department of Diagnostic Imaging, Brown University Health, Providence, Rhode Island, USA}
\affil[7]{Department of Biostatistics and Informatics, Colorado School of Public Health, Aurora, CO, USA}
\affil[8]{School of Humanities, Central South University,Changsha,Hunan,China}
\affil[9]{Department of Radiology, Second Xiangya Hospital, Central South University, Changsha, Hunan, China}
\affil[*]{H.G., C.L., Y.W.and R.W.contributed equally to this work.}
\affil[**]{Corresponding author: \texttt{harrison.bai@cuanschutz.edu}}

\begin{document}

\begin{abstract}
Ultrasound is a cornerstone of emergency and hepatobiliary imaging, yet its interpretation remains highly operator-dependent and time-sensitive. Here, we present a multitask vision–language agent (VLM) developed to assist with comprehensive right upper quadrant (RUQ) ultrasound interpretation across the full diagnostic workflow. The system was trained on a large, multi-center dataset comprising a primary cohort from Johns Hopkins Medical Institutions (9,189 cases, 594,099 images) and externally validated on cohorts from Stanford University (108 cases, 3,240 images) and a major Chinese medical center (257 cases, 3,178 images). Built on the Qwen2.5-VL-7B architecture, the agent integrates frame-level visual understanding with report-grounded language reasoning to perform three tasks: (i) classification of 18 hepatobiliary and gallbladder conditions, (ii) generation of clinically coherent diagnostic reports, and (iii) surgical decision support based on ultrasound findings and clinical data. The model achieved high diagnostic accuracy across all tasks, generated reports that were indistinguishable from expert-written versions in blinded evaluations, and demonstrated superior factual accuracy and information density on content-based metrics. The agent further identified patients requiring cholecystectomy with high precision, supporting real-time decision-making. These results highlight the potential of generalist vision–language models to improve diagnostic consistency, reporting efficiency, and surgical triage in real-world ultrasound practice.
\end{abstract}

\flushbottom
\maketitle
\thispagestyle{empty}

\section*{Introduction}

Ultrasound of the right upper quadrant (RUQ) is the preferred imaging modality for evaluating hepatobiliary disease, serving as the key tool for diagnosing and characterizing gallbladder and biliary pathology. \cite{revzin2017right} Patients with suspected findings on CT often undergo confirmatory RUQ ultrasound to assess features such as gallstones, wall thickening, and bile duct dilation, which directly guide surgical decision-making. \cite{blaivas2007diagnostic} However, RUQ ultrasound interpretation remains highly operator-dependent and limited by image noise, artifacts, and variability in acquisition and expertise. These factors can reduce diagnostic reliability and consistency across readers. \cite{kendall2001performance, gaspari2009learning} A vision–language model (VLM) designed for RUQ ultrasound could help overcome these limitations by integrating visual and clinical context to standardize interpretation, highlight key findings, and generate structured, decision-ready reports that support accurate and timely surgical management.

Early advances in AI have shown substantial promise in ultrasound interpretation across diverse applications, including the detection and classification of thyroid nodules \cite{ma2017pre, li2019diagnosis, chi2017thyroid}, breast masses \cite{byra2019breast, han2017deep, zhang2016deep}, cervical lymph nodes \cite{zhang2016application}, and liver lesions \cite{han2017deep, chaiteerakij2024artificial, akkus2019survey}. However, conventional convolutional neural network (CNN)–based systems are inherently task-specific, typically generating narrow outputs such as binary classifications or segmentation maps. This design limits their ability to integrate contextual information, perform multiple tasks, or emulate the broader clinical reasoning required for comprehensive diagnostic decision-making. \cite{hosny2018artificial}.

Recent advances in transformer-based VLMs have enabled joint learning from visual and textual modalities, offering flexible, generalizable performance across diverse medical imaging tasks. \cite{shamshad2023transformers,wu2025vision,mohsan2022vision,li2023lvit,hsu2025mri,zhao2026artificial}. In radiology, VLMs have shown strong capabilities in disease classification from chest X-rays \cite{bluethgen2025vision,zhou2025dataset}, automated report generation for CT \cite{chen2025dia,zhong2025vision,wu2025vision}, and visual question answering across multiple imaging domains \cite{bazi2023vision,alsabbagh2025minimedgpt,ayaz2024medvlm}—often performing well even in zero-shot settings where no task-specific training data are available\cite{shu2022test}. However, existing research has not yet extended these capabilities to ultrasound cine loops—dynamic, variable-length sequences that capture real-time anatomy and physiology, forming the foundation of RUQ ultrasound interpretation\cite{barr2020contrast}. Nor have current models demonstrated true multitask generalization across the full scope of ultrasound use, from image-based diagnosis to structured reporting and decision support. Addressing these gaps is crucial for developing VLMs that can operate as robust, clinically integrated assistants—capable of supporting end-to-end interpretation and guiding surgical decision-making in real-world RUQ imaging workflows. 

\begin{figure}[!htbp]
\centering
\includegraphics[width=0.75\linewidth]{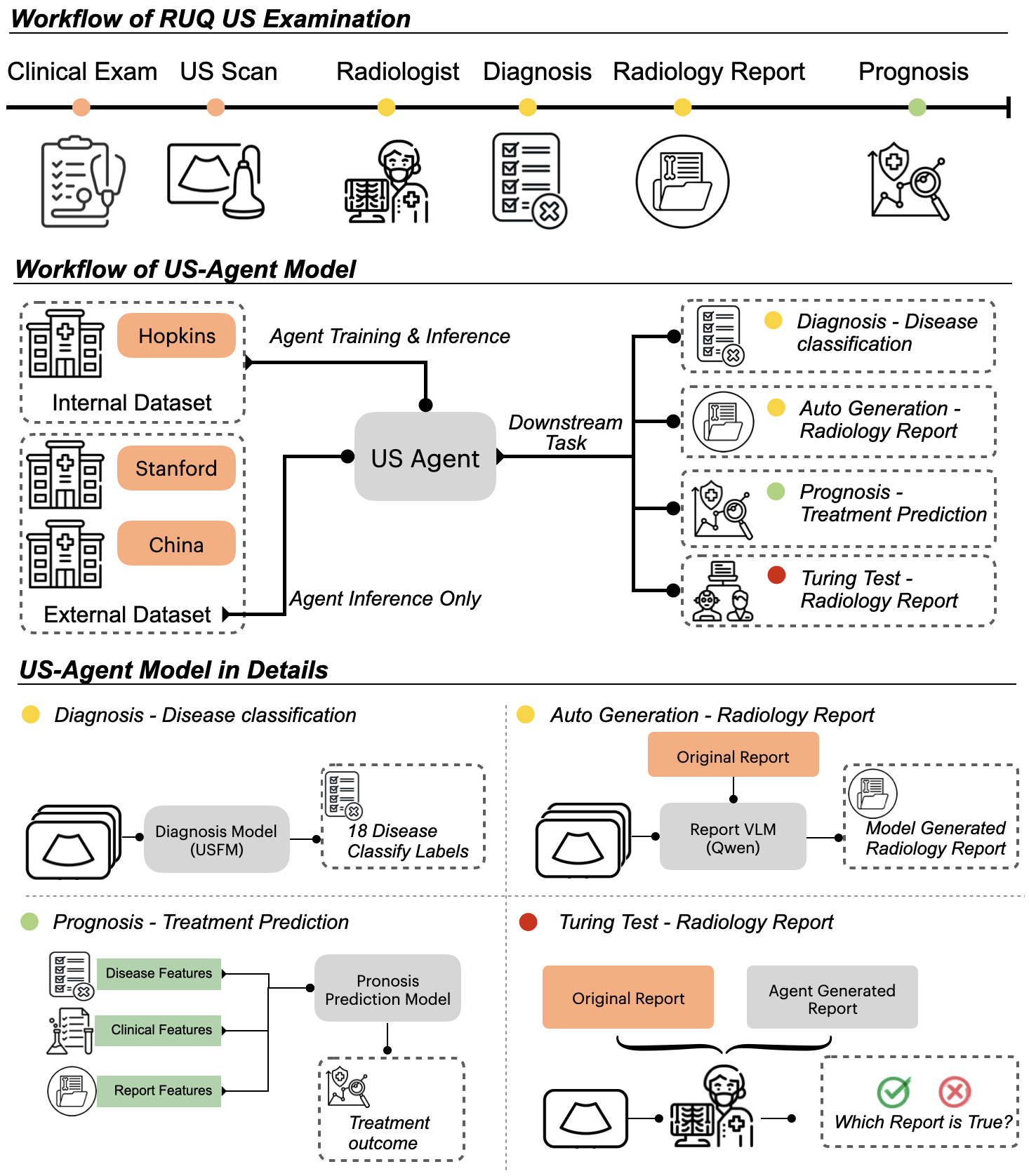}
\caption{\textbf{Overview of the right upper-quadrant (RUQ) ultrasound workflow and the proposed US-Agent model.}
\textbf{a,} Clinical workflow of RUQ ultrasound examination, illustrating the sequential steps from clinical evaluation and ultrasound acquisition to radiologist interpretation, diagnostic reporting, and surgical decision-making.
\textbf{b,} Workflow of the \textit{US-Agent} model. The internal Johns Hopkins dataset is used for both training and inference, while external datasets from Stanford and a major Chinese medical center are used for independent testing. The trained agent supports three downstream tasks: abnormality classification, automated report generation, treatment decision guidance. A radiologist Turing test was performed to assess realism and clinical accuracy.
\textbf{c,} Detailed structure of each downstream task. The abnormality classification module (USFM) predicts 18 diagnostic labels; the report generation module (Qwen VLM) produces structured radiology reports; the treatment decision module integrates image, report, disease, and clinical features to identify patients who require cholecystectomy; and the Turing test compares model-generated versus radiologist-written reports to assess realism and diagnostic fidelity.}
\label{us_framework}
\end{figure}

In this study, we introduce a VLM agent for multi-frame RUQ ultrasound interpretation, designed to support realistic, end-to-end clinical workflows in emergency care. Our work makes three key contributions. First, we curated a large-scale, multi-center dataset (n = 9,555), including a primary cohort from Johns Hopkins University and two external validation cohorts from Stanford University and a major Chinese medical center. Second, we developed a unified three-stage framework that reflects the typical RUQ ultrasound workflow: (1) Abnormality classification to automatically detect and localize pathologic findings; (2) Diagnostic report generation to produce coherent, clinically grounded narratives consistent with expert radiology standards; and (3) Treatment decision guidance to predict downstream management steps. Third, we performed blinded expert evaluations to assess the realism, diagnostic accuracy, and clinical acceptability of the model’s outputs, demonstrating its potential for integration into real-world diagnostic pipelines. Overall, this work establishes a comprehensive framework that connects clinical motivation, data scale, model design, and rigorous evaluation, paving the way for practical deployment of vision–language agents in ultrasound interpretation (Figure \ref{us_framework}).

\section*{Results}
\subsection*{Study Patients and Label Extraction}
To simulate a realistic ED workflow, we curated a multi-center dataset comprising 9,555 RUQ ultrasound text-image series. This included a primary cohort from Johns Hopkins University (9,189 cases, 594,099 images), along with two external validation cohorts from Stanford University (108 cases, 3,240 images) and a major Chinese medical center (257 cases, 3,178 images). The original ultrasound reports followed templated narrative structures with predefined sections and standardized terminology, but lacked explicit structured labels. To enable supervised learning, we employed a locally deployed Llama 3 language model \cite{grattafiori2024llama3herdmodels} to extract binary diagnostic labels across 42 categories encompassing the hepatic, biliary, renal, pancreatic, vascular, and other abdominal systems. Biliary (45\%) and hepatic (30\%) findings were the most common, followed by renal (15\%) and miscellaneous categories such as ascites and pleural effusion (10\%). In total, 37,849 positive findings were identified, with an average of 3.8 findings per patient. 

\subsection*{Abnormality Classification}
To evaluate the ability of the VLM to identify clinically relevant abnormalities on RUQ ultrasound, we fine-tuned pre-trained encoders to classify 18 diagnostic categories from multi-frame studies. These 18 categories were selected from the 42 initially extracted labels to focus on the most prevalent findings and mitigate severe class imbalance. The classification pipeline consisted of three major components. First, a feature extractor, either the Ultrasound Foundation Model (USFM), a domain-specific transformer pretrained on two million ultrasound images for segmentation, classification, and image enhancement \cite{jiao2024usfm}, or a general-purpose ResNet-50 pretrained on the ImageNet dataset. Second, an attention-based feature aggregator that adaptively weighted frame-level embeddings to generate a patient-level representation. Third, a multi-label classification head that mapped the 768-dimensional patient embedding to 18 disease probabilities using dynamically learned thresholds.

The USFM-based model achieved a macro AUROC of 0.826 (95\% CI 0.814–0.838) and a micro AUROC of 0.865 (95\% CI 0.857–0.873). At the per-disease level (Table~\ref{tab:per_disease_performance}), conditions such as right pleural effusion (AUROC = 0.939) and hepatic steatosis (AUROC = 0.912) achieved excellent performance (AUROC > 0.90). Thirteen additional conditions demonstrated good discriminative ability with AUROCs between 0.70 and 0.90. Only three conditions (gallbladder distention (0.740), coarse hepatic echotexture (0.689), and pancreas poorly visualized (0.670)) showed more limited performance. Pearson correlation analysis revealed no significant association between disease prevalence and model performance (r = -0.040 $p$ = 0.874). 

\begin{figure}[!htbp]
\centering
\includegraphics[width=0.75\linewidth]{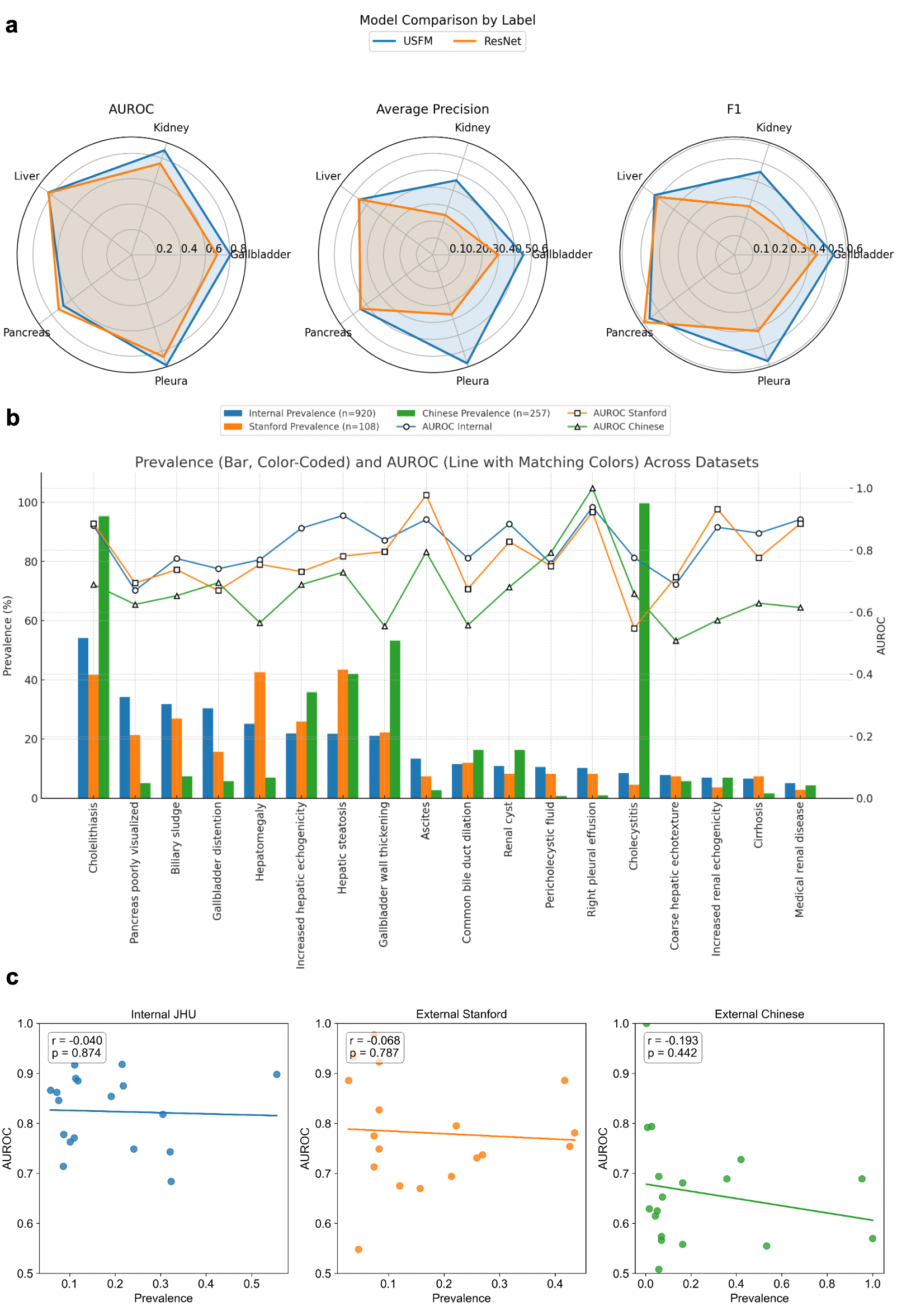}
\caption{\textbf{Performance of the automated abnormality classifier and cross-dataset generalization.} 
\textbf{a,} Per-label comparison of classification metrics between the Ultrasound Foundation Model (USFM) and the ResNet-50 baseline across representative organ systems. The USFM consistently outperformed the ImageNet-pretrained ResNet across all categories. 
\textbf{b,} Comparison of disease prevalence (bars, color-coded by dataset) and corresponding AUROC (lines of matching color) across the cohorts. Despite substantial variations in disease prevalence, model performance remained stable across datasets. 
\textbf{c,} Correlation between disease prevalence and AUROC for the internal and external validation cohorts. No statistically significant association was observed ($p > 0.05$), confirming the model’s robustness to label imbalance and institutional variability.}
\label{performance_of_classifier}
\end{figure}

External validation on two independent datasets further assessed model generalizability. The USFM-based classifier achieved a macro AUROC of 0.781 on the Stanford dataset (n = 108) and 0.667 on the Chinese external dataset (n = 257). Per-disease performance and prevalence distributions across cohorts are summarized in Figure \ref{performance_of_classifier}.

\begin{table}[htbp]
\centering
\scriptsize
\begin{tabular}{llrrrrrrllll}
\hline
 Disease & Prevalence & AUROC & Precision & Recall & F1-Score & AP & Optimal Threshold & Sensitivity & Specificity & PPV & NPV \\
\hline
Cholelithiasis & 54.1\% & 0.879 & 0.836 & 0.777 & 0.806 & 0.847 & 0.52 & 77.7\% & 89.4\% & 83.6\% & 85.2\% \\
Right pleural effusion & 10.2\% & 0.939 & 0.634 & 0.518 & 0.571 & 0.723 & 0.31 & 51.8\% & 97.1\% & 63.4\% & 95.8\% \\
Hepatic steatosis & 21.8\% & 0.912 & 0.698 & 0.635 & 0.665 & 0.781 & 0.43 & 63.5\% & 94.2\% & 69.8\% & 92.7\% \\
Ascites & 13.4\% & 0.899 & 0.612 & 0.544 & 0.577 & 0.695 & 0.28 & 54.4\% & 96.8\% & 61.2\% & 95.9\% \\
Medical renal disease & 5.1\% & 0.899 & 0.387 & 0.127 & 0.203 & 0.445 & 0.15 & 12.7\% & 99.2\% & 38.7\% & 95.8\% \\
Renal cyst & 10.9\% & 0.885 & 0.589 & 0.505 & 0.544 & 0.651 & 0.35 & 50.5\% & 96.5\% & 58.9\% & 95.4\% \\
Increased renal echogenicity & 7.0\% & 0.874 & 0.421 & 0.234 & 0.299 & 0.512 & 0.22 & 23.4\% & 98.1\% & 42.1\% & 94.8\% \\
Increased hepatic echogenicity & 21.9\% & 0.871 & 0.621 & 0.535 & 0.575 & 0.692 & 0.41 & 53.5\% & 93.8\% & 62.1\% & 91.5\% \\
Cirrhosis & 6.6\% & 0.855 & 0.498 & 0.398 & 0.442 & 0.567 & 0.25 & 39.8\% & 97.4\% & 49.8\% & 95.9\% \\
Gallbladder wall thickening & 21.1\% & 0.832 & 0.543 & 0.481 & 0.510 & 0.623 & 0.38 & 48.1\% & 92.7\% & 54.3\% & 90.8\% \\
Cholecystitis & 8.5\% & 0.776 & 0.298 & 0.156 & 0.202 & 0.389 & 0.19 & 15.6\% & 97.8\% & 29.8\% & 93.2\% \\
Common bile duct dilation & 11.5\% & 0.774 & 0.445 & 0.312 & 0.367 & 0.478 & 0.29 & 31.2\% & 95.9\% & 44.5\% & 93.8\% \\
Biliary sludge & 31.8\% & 0.773 & 0.598 & 0.489 & 0.538 & 0.634 & 0.48 & 48.9\% & 89.6\% & 59.8\% & 85.1\% \\
Hepatomegaly & 25.1\% & 0.769 & 0.521 & 0.445 & 0.480 & 0.567 & 0.42 & 44.5\% & 90.8\% & 52.1\% & 88.2\% \\
Pericholecystic fluid & 10.6\% & 0.757 & 0.198 & 0.089 & 0.131 & 0.298 & 0.16 & 8.9\% & 98.5\% & 19.8\% & 91.4\% \\
Gallbladder distention & 30.4\% & 0.740 & 0.516 & 0.398 & 0.451 & 0.542 & 0.45 & 39.8\% & 88.7\% & 51.6\% & 83.2\% \\
Coarse hepatic echotexture & 7.8\% & 0.689 & 0.298 & 0.194 & 0.239 & 0.356 & 0.21 & 19.4\% & 97.2\% & 29.8\% & 94.1\% \\
Pancreas poorly visualized & 34.2\% & 0.670 & 0.598 & 0.478 & 0.533 & 0.587 & 0.51 & 47.8\% & 83.4\% & 59.8\% & 76.8\% \\
\hline
\end{tabular}
\caption{Per-disease classification performance of the proposed abnormality classifier across 18 clinically relevant conditions. AUROC: area-under-receiver-operating-curve; AP: average precision; PPV: positive predictive value; NPV: negative predictive value }
\label{tab:per_disease_performance}
\end{table}

\subsection*{Automated Radiology Report Generation}

Radiology report generation lies at the center of the diagnostic workflow, serving as the key mechanism through which imaging findings are communicated as clinically actionable information. In the emergency department (ED), where timely and standardized interpretation is essential, report generation provides a structured synthesis of image-derived evidence that guides downstream decision-making. To assess the ability of VLMs to produce clinically meaningful ultrasound reports directly from multi-frame inputs, we trained three instruction-tuned architectures, including the M3D visual encoder paired with either LLaMA-3.1 (8B) or Phi-3-mini-4k (3.8B), and the Qwen2.5-VL (7B) model, resulting in three variants: UltrasoundLlama, UltrasoundPhi, and UltrasoundQwen. 

Traditional natural language generation (NLG) metrics are widely used to capture fluency and lexical overlap between the generated reports and the ground-truth radiology reports. As demonstrated in figure~\ref{performance_of_classifier}, all models achieved strong semantic fidelity (BERT-F1 > 0.93) on the internal JHU test set. Among them, UltrasoundQwen achieved significantly superior overall performance ($p < 0.001$; BLEU-1 = 0.5902, BLEU-4 = 0.3940, ROUGE-1 = 0.6572, ROUGE-L = 0.5527, METEOR = 0.5506). Between the M3D-based variants, UltrasoundPhi achieved significantly higher BLEU-1 scores ($p < 0.05$), indicating better short-form lexical precision, whereas UltrasoundLLaMA outperformed in ROUGE-L ($p < 0.05$) and METEOR ($p < 0.01$), reflecting stronger phrase-level recall and synonym alignment.

To assess generalizability, we further benchmarked the models on two external datasets from Stanford University and a major Chinese medical center. Consistent with internal findings, UltrasoundQwen maintained significantly higher performance than both UltrasoundLLaMA and UltrasoundPhi across multiple lexical metrics (Stanford: BLEU-1 and ROUGE-L; Chinese: all metrics). Notably, UltrasoundPhi surpassed UltrasoundLLaMA across all metrics ($p < 0.05$) on both external cohorts, contrasting with their mixed internal results. These findings suggest superior cross-institutional generalizability of the Qwen and M3D-Phi configurations compared with the M3D-LLaMA variant. Full quantitative comparisons are presented in Table \ref{tab:report_generation}.

Since NLG metrics often fail to reflect the clinical accuracy and informational density required for diagnostic use, we further evaluated all models using two clinically oriented frameworks: DocLens\cite{xie2023doclens}, an LLM-as-a-judge tool assessing factual recall, and FORTE\cite{li2025towards}, which quantifies the inclusion of key clinical information across four dimensions (Degree, Landmark, Feature, Impression). As summarized in figure~\ref{performance_of_classifier}, UltrasoundQwen demonstrated superior performance on both conventional and clinically focused metrics. On the internal JHU dataset, it achieved a DocLens recall of 0.6094 and a FORTE average F1-score of 0.6557, significantly outperforming UltrasoundLLaMA and Ultrasound Phi. The advantage of UltrasoundQwen was preserved across both external cohorts. In the Chinese dataset, it achieved a FORTE average of 0.6181, a substantial improvement over its competitors, highlighting its strong cross-lingual and cross-institutional generalization. All comparative results across metrics and institutions are visualized in Table~\ref{tab:clinical_report_gen}.

\begin{figure}[!htbp]
\centering
\includegraphics[width=0.75\textwidth]{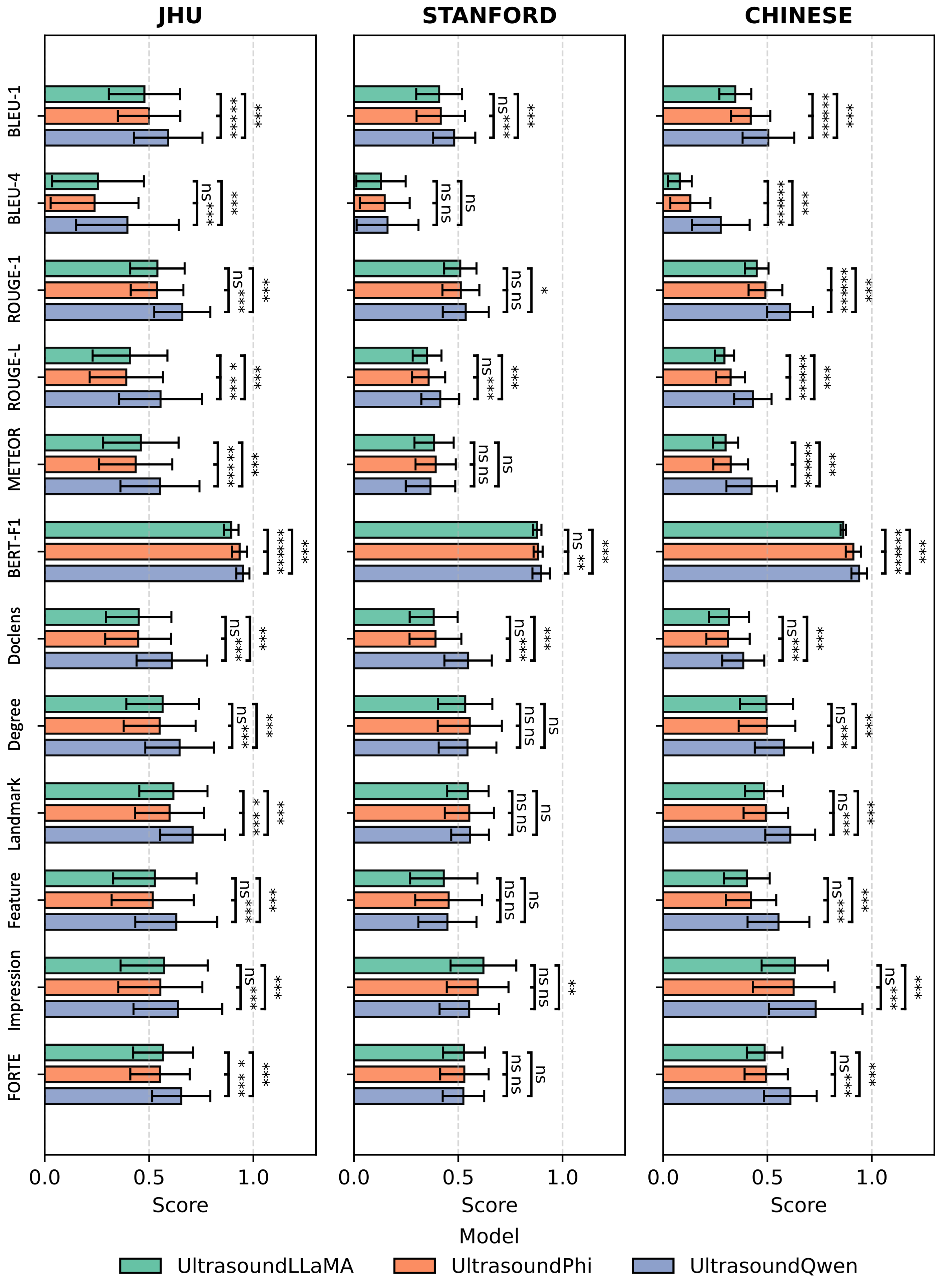}
\caption{\textbf{Cross-institutional comparison of automated report generation performance.}
Bar plots summarizing the performance of UltrasoundLLama, UltrasoundPhi, and UltrasoundQwen across the internal JHU dataset and two external validation cohorts from Stanford University and a major Chinese medical center. Each metric represents the mean of three independent runs, with standard deviations shown as error bars. Text similarity metrics (BLEU, ROUGE, METEOR, BERT-F1) capture linguistic and semantic fidelity, whereas clinical realism metrics (DocLens\cite{xie2023doclens}, and the FORTE\cite{li2025towards} categories) assess factual accuracy and structural correspondence to expert-written reports. UltrasoundQwen consistently achieved the highest scores across all datasets. Statistical significance is denoted by asterisks (* $p<0.05$, ** $p<0.01$, *** $p<0.001$).}
\label{fig:report_generation_comparison}
\end{figure}

\subsection*{Cholecystectomy treatment recommendation}
To assess the agent's ability to support downstream clinical decision-making, we evaluated whether a large language model (LLM) could predict the need for cholecystectomy using multimodal clinical inputs. We implemented a Llama3 model and focused the analysis on a cohort of 2,170 patients with confirmed cholecystitis, where the surgical decision is clinically critical.

We first established a benchmark using ground-truth data that included patient demographics, ICD codes, and the original radiologist reports. As shown in Table~\ref{tab:cholecystectomy_prediction}, the benchmark model achieved its highest performance when all input features were combined, yielding an \text{accuracy of 0.843} and a \text{recall of 0.665}. Ablation analysis revealed that each modality contributed a meaningful signal to the model’s performance, with the combination of ICD codes and narrative reports providing the greatest improvement in recall compared with patient variables alone.

We then tested whether our AI agent could replicate this predictive capability using only machine-generated inputs. To do so, we replaced the human-written radiology reports with those generated by our VLM and repeated the analysis. The results demonstrated that the VLM-generated reports served as an effective proxy for ground-truth narratives. The model incorporating these generated reports, along with demographic and ICD features, achieved an \text{accuracy of 0.718} and a \text{recall of 0.602}. Although a performance gap remains compared with the benchmark, the model retained strong recall, indicating that the VLM successfully captured and conveyed the essential clinical information required for this complex downstream prediction task.

\begin{table}[htbp]
\centering
\caption{\textbf{Performance of LLM-based cholecystectomy treatment recommendation model.} Comparison of model accuracy and recall using original human-written radiology reports versus vision–language model (VLM)–generated reports as input. Each configuration combines different data modalities, including patient demographics (“Variables”), ICD diagnostic codes, and either human or generated reports.}
\label{tab:cholecystectomy_prediction}
\begin{tabular}{l cc cc}
\toprule
& \multicolumn{2}{c}{\textbf{Original Report}} & \multicolumn{2}{c}{\textbf{Generated Report}} \\
\cmidrule(lr){2-3} \cmidrule(lr){4-5}
\textbf{Input Features} & \textbf{Accuracy} & \textbf{Recall} & \textbf{Accuracy} & \textbf{Recall} \\
\midrule
Variables & 0.869 & 0.126 & 0.868 & 0.128 \\
Variables + ICD & 0.890 & 0.515 & 0.814 & 0.500 \\
Variables + Report & 0.787 & 0.431 & 0.794 & 0.438 \\
Variables + ICD + Report & 0.843 & 0.665 & 0.718 & 0.602 \\
\bottomrule
\end{tabular}
\end{table}

\subsection*{Expert Evaluation of the VLM Agent}
Blinded Turing tests were conducted to assess the ability of radiologists to distinguish between AI-generated and human-written ultrasound reports. Radiologists identified unedited AI-generated reports with an average accuracy of 0.85 (95\% CI: 0.76-0.93), while accuracy decreased to 0.46 (95\% CI: 0.30-0.62) when evaluating post-edited AI-generated reports ($p < 0.05$). The average evaluation time was 1.18 (95\% CI: 0.76-1.32) minutes per case for unedited AI reports and 1.82 (95\% CI: 1.42-2.01) minutes per case for post-edited AI reports ($p < 0.05$). For the combined set of unedited and edited AI reports, the average accuracy of identifying the human-written report as ground truth was 0.61 (95\% CI: 0.34-0.98) when only text was provided and 0.69 (95\% CI: 0.47-0.88) when both text and corresponding ultrasound images were available ($p > 0.05$). When stratified by experience level, expert radiologists achieved accuracies of 0.56 (95\% CI: 0.40–0.70) (text-only) and 0.69 (95\% CI: 0.604–0.850) (text plus image), whereas trainees achieved 0.68 (95\% CI: 0.44–0.72) (text-only) and 0.64 (95\% CI: 0.34–0.70) (text plus image)($p > 0.05$). 

\begin{figure}[!htbp]
\centering
\includegraphics[width=0.99\linewidth]{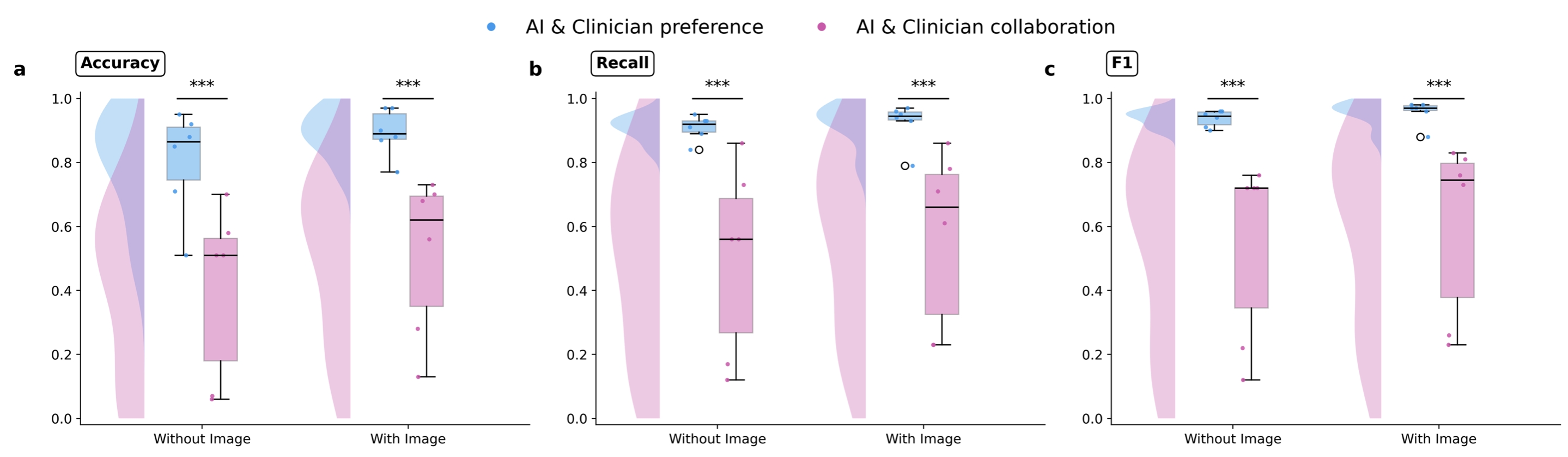}
\caption{\textbf{Turing test evaluation Results.} 
Boxplots show accuracy (a), recall (b), and F1 score (c) for radiologist judgments in both settings (blue = AI versus Clinician preference, pink = AI + Clinician collaboration) under conditions with and without ultrasound images. Collaboration and image access significantly improved performance across all metrics (*** $p < 0.001$).}
\label{turing_test_0}
\end{figure}

\begin{table}[ht]
\centering
\caption{\textbf{Turing Test Performance in Distinguishing AI from Human-Generated Reports.} Accuracy represents the proportion of cases in which radiologists correctly identified the human-written report as the ground truth. Lower accuracy indicates greater indistinguishability between AI-generated and expert-written reports.}
\label{tab:turing_test_summary}
\begin{tabular}{llc}
\toprule
\textbf{Comparison} & \textbf{Condition} & \textbf{Accuracy} \\
\midrule
\multicolumn{3}{l}{\textbf{Report Type}} \\
& Initial AI Report & 0.85 \\
& Post-Edited AI Report & 0.46 \\
\midrule
\multicolumn{3}{l}{\textbf{Imaging Context}} \\
& Without Images & 0.61 \\
& With Images & 0.69 \\
\midrule
\multicolumn{3}{l}{\textbf{Expertise Level (With vs. Without Image)}} \\
Expert & Without Image & 0.56 \\
Expert & With Image & 0.69 \\
Trainee & Without Image & 0.68 \\
Trainee & With Image & 0.64 \\
\bottomrule
\end{tabular}
\end{table}
\clearpage 
\section*{Discussion}
In this study, we developed and validated a vision–language agent framework for automated interpretation of multi-frame RUQ ultrasound, designed to emulate the complete clinical workflow from image analysis to decision support. Our results show that our multi-task system can accurately perform three core tasks traditionally requiring physician expertise -abnormality classification, diagnostic report generation, and treatment recommendation- thereby establishing a foundation for practical physician–AI integration in emergency care.

Our agent trained on the Ultrasound Foundation Model (USFM) demonstrated clear superiority over a generic ResNet-50 backbone. This highlights the value of large-scale, domain-specific pre-training for medical imaging. Similar to recent ultrasound foundation and vision–language systems such as USFM \cite{jiao2024usfm}, URFM \cite{kang2025urfm}, EchoVLM \cite{she2025echovlm}, and UltraSAM \cite{meyer2025ultrasam}, our results confirm that representation learning on ultrasound data yields substantial gains that cannot be matched by generic visual encoders. These studies collectively show that domain-specialized visual encoders not only enhance classification but also support report generation and downstream reasoning tasks. Nevertheless, the observed performance gap across institutions, particularly on the Chinese external cohort, underscores the persistent challenge of cross-site generalization, consistent with prior reports on ultrasound domain shift and cross-site transferability \cite{rajagopal2023federated, liu2025visual}. Differences in patient populations, scanning conventions, and device vendors remain key sources of variation. Future work should incorporate domain-adaptation and adversarial learning strategies \cite{ganin2016domain} as well as federated or continual learning pipelines \cite{rajagopal2023federated} to promote robustness across global clinical settings.

The diagnostic reports generated by UltrasoundLLaMA, UltrasoundPhi, and UltrasoundQwen represent a significant step beyond conventional text-similarity metrics commonly used in medical natural language generation. Earlier report-generation systems—such as LLAUS \cite{guo2025llaus}, EchoVLM \cite{she2025echovlm}, and USFM \cite{jiao2024usfm}—have advanced the integration of visual reasoning with narrative synthesis, yet their evaluation has largely depended on surface-level fluency measures (e.g., BLEU, ROUGE). In contrast, by applying clinically weighted evaluation frameworks such as FORTE \cite{li2025towards} and DocLens\cite{xie2023doclens}, we assessed factual precision and informational completeness rather than linguistic overlap alone. The strong performance of the UltrasoundQwen (FORTE F1 = 0.66, DocLens = 0.61) model demonstrates that its generated narratives accurately capture key diagnostic content, characterize pathologic findings, identify anatomic landmarks, and maintain impression consistency—approaching the standard of expert radiology reporting.

Importantly, the output from Ultrasound Qwen encodes clinically decisive information to be actionable in our large-language-model classifier to predict the need for cholecystectomy with high recall (0.602). This extends the utility of vision–language systems from description to inference, paralleling recent work that couples report generation with clinical decision support and outcome prediction \cite{zhong2025vision,ZHONG2026103786}. The ablation results further reveal that report generation is not merely cosmetic but transforms raw visual data into an interpretable and clinically relevant representation suitable for downstream reasoning, an approach aligned with emerging paradigms of reasoning-centric VLMs in medicine \cite{liu2025visual,wang2024enhancing}.

The blinded expert evaluation illustrates how human oversight enhances model credibility. Although unedited AI reports were generally identifiable, minimal post-editing (68.4 s $\pm$ 21.0 s on average per report in this paper) rendered them statistically indistinguishable from human-written reports, suggesting a realistic “AI-copilot” paradigm in which radiologists verify and refine AI-drafted outputs rather than authoring from scratch. 
There are several possible reasons why clinician–AI collaboration (“AI-copilot” paradigm) can be effective. First, even with a baseline level of inter-rater variability in preference and error correction, AI assistance provides a consistent reference that helps reduce subjective differences among clinicians. Second, collaboration between clinicians and AI can enhance both efficiency and diagnostic precision when the AI system provides complementary insights rather than competing judgments. Developing robust strategies for when and how to present AI-generated drafts is, therefore, crucial to maximizing the advantages of assistance.
Third, it is plausible that revising an AI-generated report can substantially reduce reporting time compared to writing from scratch. Although this was not explicitly quantified in our study, time savings represent an important direction for future work, especially given that reporting speed depends on clinical context, expertise, and case complexity.
Such collaboration could reduce documentation burden while preserving professional accountability. The differential performance between expert and trainee readers—experts benefiting more from image context, suggesting that clinical experience remains essential for adjudicating subtle findings, whereas AI contributes scale, speed, and consistency.

Several limitations warrant acknowledgment. The retrospective design and label extraction using a large language model introduce weak supervision that may omit fine-grained diagnostic nuance. The cholecystectomy prediction model was trained on a limited subset of patients, and broader, prospective validation is needed to confirm generalizability across diverse institutions and clinical scenarios.

Overall, this work establishes a comprehensive, end-to-end framework linking image perception, linguistic reasoning, and therapeutic guidance within a single multimodal agent. By coupling scalable foundation models with expert oversight, such systems could accelerate radiologic reporting, standardize interpretation quality, and extend diagnostic capacity to resource-limited settings. The next critical step is prospective clinical evaluation to determine how these physician–AI partnerships influence diagnostic accuracy, workflow efficiency, and ultimately patient outcomes in real-world emergency care.
\clearpage 
\section*{Methods}
\subsection*{Study Design and Data Acquisition}
\subsubsection*{Multi-center cohort description}
This retrospective study utilized a multi-center dataset comprising 9,555 RUQ ultrasound text–image series. The primary cohort was obtained from the Johns Hopkins University and included 9,189 unique patient cases after quality filtering. Data collection for this cohort spanned from July 31, 2016, to December 31, 2023. Patients were included if they underwent a clinically indicated RUQ ultrasound examination. Exclusion criteria included incomplete data or technically limited images that precluded diagnostic interpretation. For external validation, two additional cohorts were incorporated. The first external cohort consisted of 108 cases (3,240 images) from Stanford University, and the second external cohort comprised 257 cases (3,178 images) from Second Xiangya Hospital in Changsha, China. All study protocols were reviewed and approved by the Institutional Review Boards of the Johns Hopkins University, Stanford University, and the participating Chinese medical center. Given the retrospective design and the use of fully de-identified data, the requirement for informed consent was waived by all institutional ethics committees.

\subsubsection*{Automated Report Processing and Label Extraction with a Local LLM}
To generate structured labels for model training and evaluation, we developed an automated natural language processing pipeline to analyze all 9,555 free-text RUQ ultrasound reports corresponding to the image datasets. The pipeline employed a locally deployed Llama 3 70B Instruct model \cite{Dubey2024Llama3} to parse the narrative Findings and Impression sections of each report. Using optimized prompt engineering, the system extracted and standardized diagnostic information into 42 predefined medical categories spanning the hepatic, biliary, renal, pancreatic, and vascular systems. This automated extraction process identified a total of 37,849 positive findings, corresponding to an average of 3.8 findings per patient across the entire cohort. While this process identified 42 distinct categories, a preliminary analysis revealed that many had a very low prevalence. To create a robust dataset for the subsequent classification task and mitigate severe class imbalance, we selected only the 18 labels with a prevalence greater than 5\% for model training. 

\subsection*{VLM Agent Framework}
\subsubsection*{Image preprocessing and standardization}
Our dataset is composed of clinical ultrasound cine loops (video sequences), which are inherently variable in length, ranging from under 20 to over 200 frames. To create a uniform input for our deep learning model, we developed a preprocessing pipeline to standardize every cine loop to a fixed length of 32 frames.

The core of our method is uniform temporal sampling. For longer videos, we deterministically select 32 frames evenly spaced across the entire duration of the scan. For the few videos shorter than 32 frames, we perform cyclical temporal padding by repeating the sequence to meet the required length. This two-pronged approach ensures that the standardized input is a faithful temporal representation of the original scan, capturing information from its beginning, middle, and end phases. This process preserves the temporal dynamics critical for our analysis while creating a fixed-size input required by the model. A detailed, step-by-step description of this standardization algorithm is provided in the Supplementary Material (Section S1).

Following frame selection, an automated workflow was used to remove artifacts and standardize the image inputs. This included isolating the diagnostically relevant fan-shaped scanning sector and eliminating non-imaging elements such as embedded text overlays, ECG tracings, and display markers. The selected frames were then resized to a uniform resolution of 224 × 224 pixels using bicubic interpolation, and pixel intensities were normalized to a standardized range of [0.0, 1.0] to ensure consistent input for downstream model training.

\subsubsection*{Multi-task agent architecture}
To emulate the clinical workflow of ultrasound interpretation, we developed an end-to-end, multi-task agent framework comprising three sequential stages: (1) Abnormality Classification, (2) Automated Report Generation, and (3) Treatment Recommendation. The architecture was designed so that the output from each stage serves as a conditional input for the subsequent one, enabling progressive information transfer across the diagnostic and decision-making pipeline. In the Abnormality Classification stage, a multi-label classification model processes the 32 standardized ultrasound frames to generate patient-level predictions for 18 clinically relevant pathologies. The model produces both categorical disease labels and 768-dimensional image embeddings, which are passed to the next stage. In the Automated Report Generation stage, a VLM generates a structured diagnostic report conditioned on the visual features from the ultrasound frames and the structured predictions from the classification model. This stage links quantitative image interpretation with natural-language narrative synthesis. In the final Treatment Recommendation stage, a predictive model for cholecystectomy integrates multi-modal information—combining the image embeddings from the classifier with text embeddings derived from either human-written or VLM-generated reports—to estimate the likelihood of surgical intervention.

\subsection*{Abnormality Classification Module}
\subsubsection*{Model architecture and training}
To enable automated abnormality classification, we developed a multi-component disease classification module consisting of three core elements. The first component was a visual feature extractor, implemented using the Ultrasound Foundation Model (USFM), a Vision Transformer (ViT-B/16) backbone pretrained on a large-scale, domain-specific ultrasound dataset. This backbone processed standardized input images of 224 × 224 pixels (patch size 16 × 16) and generated 768-dimensional feature embeddings for each of the 32 frames per patient. The second component was an attention-based feature aggregator designed to handle the variable number of frames across patient studies. This module learned adaptive weighting of frame-level embeddings to produce a single consolidated 768-dimensional patient-level representation.The final component was a multi-label classification head, consisting of a dropout layer (rate = 0.1) followed by a linear layer that mapped the 768-dimensional feature vector to 18 disease probabilities. Dynamically learned thresholds were applied to determine per-disease decision boundaries.

To address severe class imbalance in the training dataset, we employed Asymmetric Loss (ASL) as the primary objective function. Optimization was performed using AdamW (initial learning rate = 1 $\times$ $10^{-4}$) with a ReduceLROnPlateau scheduler that reduced the learning rate by a factor of 0.5 after 10 epochs without improvement in validation loss. Additional stabilization measures included gradient clipping (maximum norm = 5.0) and Gradient Surgery, which projected conflicting task gradients orthogonally to prevent inter-task interference. Data augmentation was implemented using Multi-Label Mixup with a Beta distribution parameter ($\alpha$ = 0.4) applied at a 50\% probability. To further mitigate class imbalance, we applied adaptive sample reweighting, assigning double weight to hard samples (top 30\% by loss) based on an exponential moving average ($\beta$ = 0.9) of per-sample losses. Class weights were also incorporated, scaled inversely to class frequency, and clamped between 1.0 and 10.0. A 5-epoch warm-up phase preceded activation of the full training strategy.

\subsubsection*{Model evaluation}
The performance of the final classifier was evaluated on a held-out internal test set of 920 patient studies from the primary Johns Hopkins cohort. Evaluation was conducted across 18 disease categories with a prevalence greater than 5\% in the training data.
To assess generalizability, the model was further evaluated on two external validation cohorts. The first included 108 cases from Stanford University, and the second comprised 257 cases from the Second Xiangya Hospital in Changsha, China. For both external datasets, performance was reported for the same 18 disease categories that could be directly mapped to the labels in the primary cohort.

\subsubsection*{Ablation study design}
To validate the contribution of the domain-specific USFM backbone, a comparative ablation study was conducted. A baseline model was constructed using a general-purpose ResNet-50 architecture, pre-trained on the ImageNet dataset, as the visual feature extractor. All other components of the framework, including the attention-based aggregator and the multi-label classification head, remained identical to the primary model. This baseline model was trained and evaluated under the exact same conditions, utilizing the same 8,269-sample training set, loss function, optimizer, learning rate schedule, and all other specified hyperparameters to ensure a direct and fair comparison of the backbones' performance.

\subsection*{Automated Report Generation Module}
\subsubsection*{Vision-language model selection and fine-tuning}
To enable multimodal report generation, we fine-tuned a suite of VLMs for the automated ultrasound reporting task. The candidate architectures included Qwen2.5-VL-7B, a 7-billion-parameter model with a Qwen2-VL \cite{bai2025qwen25vltechnicalreport} Vision Transformer backbone; M3D \cite{bai2024m3dadvancing3dmedical}
, a 3D medical VLM evaluated with two language backbones, the Phi-3-mini-4k (4B) \cite{abdin2024phi3technicalreporthighly} and Llama-3.1-8B (8B) models \cite{Dubey2024Llama3}; and Otter \cite{li2025ottermultimodalmodelincontext}, an open-source Flamingo-style model, which served as a baseline. All models were fine-tuned on the internal training dataset (n = 8,269) using a mixed-supervision, instruction-tuning strategy.

The training data were organized in a conversational format with three components:
(1) the full Findings section from the original human-written report, serving as the ground-truth response to a general instruction prompt to “generate a diagnostic report”; (2) a series of binary (Yes/No) visual question–answer pairs derived from the 13 most common GPT-extracted disease labels, designed to reinforce diagnostic accuracy; and (3) the complete set of 32 standardized ultrasound frames for each patient. For Qwen2.5-VL-7B, fine-tuning was performed using Low-Rank Adaptation (LoRA) with a rank of 16 and an alpha value of 32. The model was trained for 3 epochs with a learning rate of 1 $\times$ $10^{-4}$ and an effective batch size of 8 (1 sample per device, 8 gradient accumulation steps), using bfloat16 precision. The M3D models underwent a two-phase training pipeline consisting of a vision–language alignment pretraining phase followed by a visual instruction-tuning phase, also using LoRA (rank = 16, alpha = 32) with a learning rate of 5 $\times$ $10^{-5}$ for 5 epochs. All training was conducted on a NVIDIA DGX workstation equipped with four A100 GPUs.

\subsubsection*{Report Quality Assessment with Traditional, Clinical, and AI-driven Metrics}

The quality of the generated reports was assessed across the internal held-out test set from Johns Hopkins (n=920) and the two external validation cohorts from Stanford University (n=108) and a major Chinese medical center (n=257). To quantify report quality, we employed a multi-faceted evaluation strategy combining traditional natural language generation (NLG) metrics with two advanced, clinically-oriented frameworks designed to overcome the limitations of surface-level text similarity.

\paragraph{Traditional NLG Metrics}
To assess linguistic fluency and content overlap with ground-truth reports, we used a standard panel of metrics:
\begin{itemize}
    \item \textbf{ROUGE-1} (Recall-Oriented Understudy for Gisting Evaluation)~\cite{lin2004rouge}: Measured the recall of unigrams.
    \item \textbf{METEOR} (Metric for Evaluation of Translation with Explicit ORdering)~\cite{banerjee2005meteor}: Evaluated alignment based on exact unigrams, stemmed tokens, and synonyms.
    \item \textbf{BLEU-1} and \textbf{BLEU-4} (Bilingual Evaluation Understudy)~\cite{papineni2002bleu}: Measured the precision of 1-gram and 4-gram matches to assess grammatical structure.
    \item \textbf{BERT-F1}~\cite{zhang2020bertscore}: Computed F1 scores using contextualized embeddings to evaluate semantic similarity.
\end{itemize}

\paragraph{Clinically-Oriented Evaluation (FORTE)}
To directly measure the clinical value of generated reports, we implemented the Feature-Oriented Radiology Task Evaluation (FORTE) framework \cite{li2025towards}. FORTE deconstructs a report's clinical essence into four key components, each governed by a dedicated keyword lexicon. Following Li et al. \cite{li2025towards}, we curated the RUQ ultrasound keyword bank, where each category-specific lexicon encompasses both primary terms and clinically equivalent synonyms to capture lexical variability and differences in reporting style.
\begin{itemize}
    \item \textbf{Degree:} Describes the size, scope, and intensity of findings (e.g., \textit{dilation/ distended/ distension/ enlarged/ prominent}, \textit{extensive}, \textit{absent/ negative/ no/ not}, \textit{limited/ poorly/ suboptimally}).
    \item \textbf{Landmark:} Pinpoints the anatomical location of findings (e.g., \textit{pancreas/ pancreatic}, \textit{kidney/ renal/ urinary}, \textit{interpolar/ mid/ midportion}).
    \item \textbf{Feature:} Details the primary imaging characteristics of the findings (e.g., \textit{lesion/ mass/ masses}, \textit{size/ volume}, \textit{coarse/ coarsened/ heterogenous}).
    \item \textbf{Impression:} Captures the final diagnostic conclusions (e.g., \textit{cholelithiasis}, \textit{hepatic steatosis}, , \textit{cirrhosis}).
\end{itemize}
For evaluation, keywords from each category were extracted from both the AI-generated and ground-truth reports. By comparing the keyword matches, an F1-score was calculated for each dimension, and a final average F1-score quantified the report's overall clinical information density and accuracy. (For the keyword details, please see Supplementary Table~\ref{tab:ruq_forte_keywords})

\paragraph{AI-driven Factual Accuracy Evaluation (LLM-as-a-judge)}
To simulate an expert review and validate the factual accuracy of our VLM's outputs, we utilized an \textbf{LLM-as-a-judge} approach with the \textbf{DocLens} \cite{xie2023doclens} tool on Llama3 local LLM environment. This automated process unfolds in three steps:
\begin{enumerate}
    \item \textbf{Extract Claims:} DocLens parses the AI-generated report and extracts each distinct medical assertion.
    \item \textbf{Verify Presence:} It then cross-references each claim against the ground-truth human report to determine if it is factually supported.
    \item \textbf{Compute Score:} A final precision score is calculated based on the proportion of verified claims among all extracted claims. This score directly measures the factual reliability of the generated report, penalizing unsupported or hallucinatory statements.
\end{enumerate}

\subsection*{Treatment Recommendation for Cholecystectomy}
\subsubsection*{Patient cohort definition} 
To develop a clinically relevant prediction model for cholecystectomy, we extracted a subset of patients with a confirmed diagnosis of cholecystitis from the primary dataset using corresponding ICD diagnosis codes. The final cohort (n = 2,170) was randomly partitioned into a training set (n = 1,736; 80\%) and a held-out test set (n = 434; 20\%). This filtering process improved class balance from less than 1\% in the overall cohort to approximately 15\% in the cholecystitis-specific subset. The ground-truth outcome—receipt of cholecystectomy—was determined by identifying the corresponding procedural codes in the patients’ electronic health records.

\subsubsection*{Predictive model development and optimization}
To predict the need for cholecystectomy, we implemented a Llama 3 large language model (LLM) in a few-shot classification setting. The task was formulated as a binary classification problem, prompting the model to output “Yes” or “No” regarding the necessity of surgical intervention. Each input prompt included combinations of engineered features serialized into a structured textual format. A full prompt consisted of patient demographics, relevant ICD codes, and the full text of either the human-written or VLM-generated radiology report. Model performance was evaluated on the held-out test set using accuracy and recall as primary metrics. The evaluation emphasized the ability of VLM-generated reports to serve as a viable substitute for human-written reports in downstream decision-making tasks.

\subsubsection*{Ablation study}
To assess the relative contribution of each data modality to the predictive performance, a systematic ablation study was conducted. The model was evaluated using four configurations of input features: (1) patient variables alone, (2) variables plus ICD codes, (3) variables plus the radiology report text, and (4) all three feature types combined. Each configuration was tested using both original human-written and VLM-generated reports, enabling direct comparison of model performance across feature subsets and report sources. This analysis quantified the incremental predictive value contributed by each data modality and the degree to which AI-generated reports captured the clinical information necessary for surgical decision support

\subsection*{Expert Evaluation of the VLM Agent and Physician-AI Collaboration}
\subsubsection*{Blinded reader study design}
To evaluate the clinical acceptability and quality of AI-generated reports—and to explore potential collaboration between physicians and AI systems—we conducted a blinded reader study modeled after a Turing test framework \cite{tanno2025collaboration}. The study included two independent reader groups representing different levels of clinical experience: one group evaluated the initial AI-generated reports, and a separate group evaluated the post-edited versions. For the initial evaluation, a set of 60 non-overlapping patient cases was randomly selected from the held-out test set. Each case included two reports presented side by side in randomized order: (1) the unedited AI-generated report produced by the Qwen2.5-VL-7B model and (2) the ground-truth human-written report authored by a radiologist. The evaluation followed a two-stage protocol. In the first stage, readers reviewed the text-only reports and were asked to identify which of the two had been generated by AI. In the second stage, readers were provided with the full ultrasound image series for each case and asked to reassess their decision. This two-step process enabled evaluation of both textual plausibility and image-grounded clinical reasoning.

\begin{figure}[!htbp]
\centering
\includegraphics[width=0.85\linewidth]{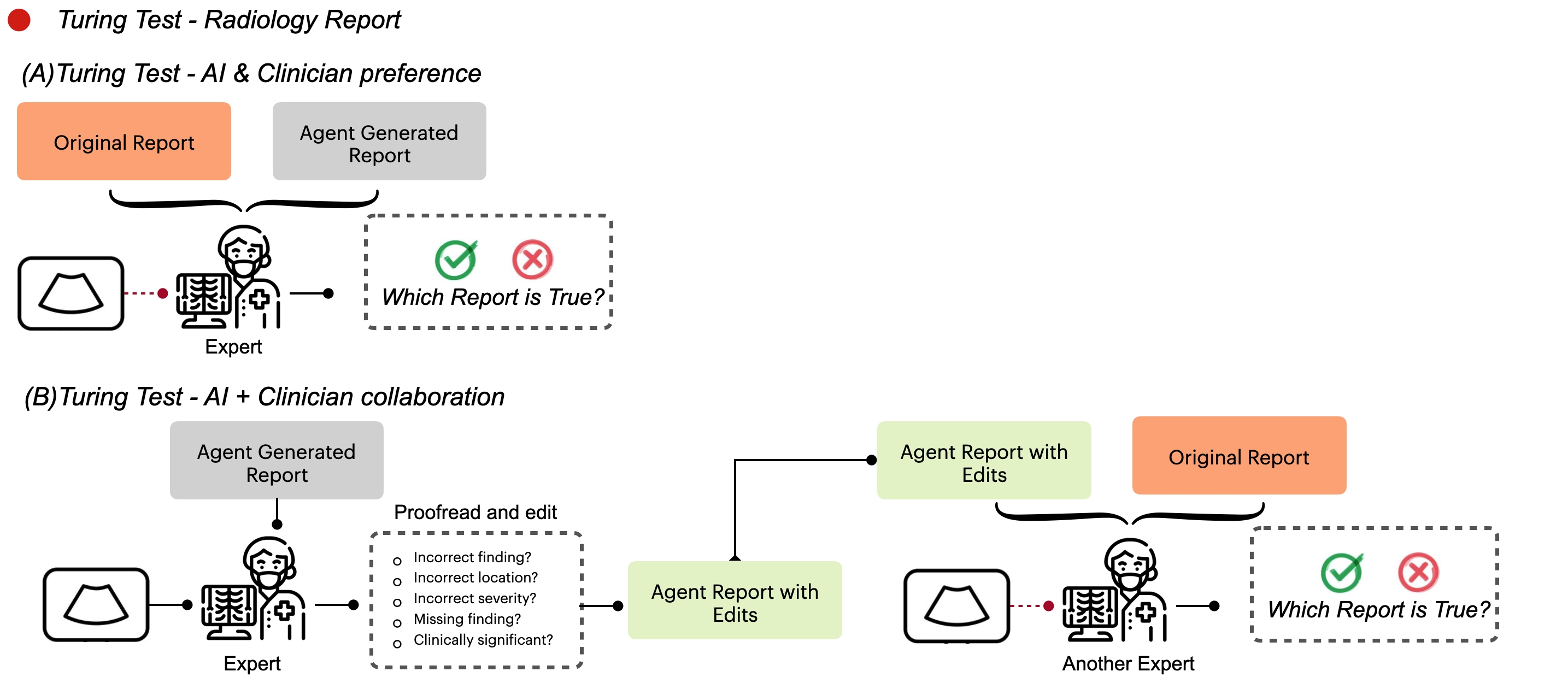}
\caption{\textbf{Turing test evaluation of AI-generated radiology reports.} 
\textbf{a} | \textit{AI versus Clinician Preference.} Radiologists were presented with ultrasound images alongside two reports—the original human-written version and the AI-generated version—and asked to identify which they believed to be the authentic human report. 
\textbf{b} | \textit{AI + Clinician Collaboration.} Radiologists proofread and post-edited AI-generated reports, correcting factual, locational, and severity errors and adding missing or clinically significant findings. The edited reports were then compared with the originals by an independent expert to evaluate indistinguishability and clinical accuracy.}
\label{turing_test_1}
\end{figure}

\subsubsection*{Evaluation of post-edited reports}
To assess whether human refinement could improve the perceived quality and realism of AI-generated reports, a second blinded reader study was conducted. For this evaluation, the AI-generated reports from a different subset of cases were manually post-edited by a board-certified radiologist with 7 years of post-fellowship experience (HXB) to correct factual inaccuracies, refine phrasing, and add clinically relevant details. A new group of six radiologists then compared these post-edited AI reports with the original human-written reports for the same cases. The evaluation protocol mirrored that of the first study, consisting of two stages: initial review of text-only reports, followed by re-evaluation after viewing the corresponding ultrasound image series. This design enabled a direct comparison of indistinguishability between the unedited and post-edited AI reports, quantifying the impact of human–AI collaboration on report realism and clinical acceptability

\subsubsection*{Statistical analysis}
Normally distributed variables (e.g., report performance or accuracy) were compared using two-tailed paired t-tests, while non-normally distributed data were analyzed with the Wilcoxon signed-rank test. Between-group comparisons (e.g., expert vs. trainee, text-only vs. text + image) used independent t-tests as appropriate. 95\% confidence intervals (CIs) were estimated via bootstrap resampling. All tests were two-sided, with p < 0.05 considered statistically significant.

\section*{Code Availability}
The code developed for this study, including the VLM framework and analysis scripts, is publicly available on GitHub at \url{https://github.com/Mikeghm/ruq-ultrasound-vlm}.

\clearpage 
\bibliography{sample}

\clearpage 
\section*{Supplementary Materials} 
\setcounter{figure}{0} 
\setcounter{table}{0} 
\renewcommand{\thetable}{S\arabic{table}} 

\begin{figure}[!htbp]
\centering
\includegraphics[width=0.95\linewidth]{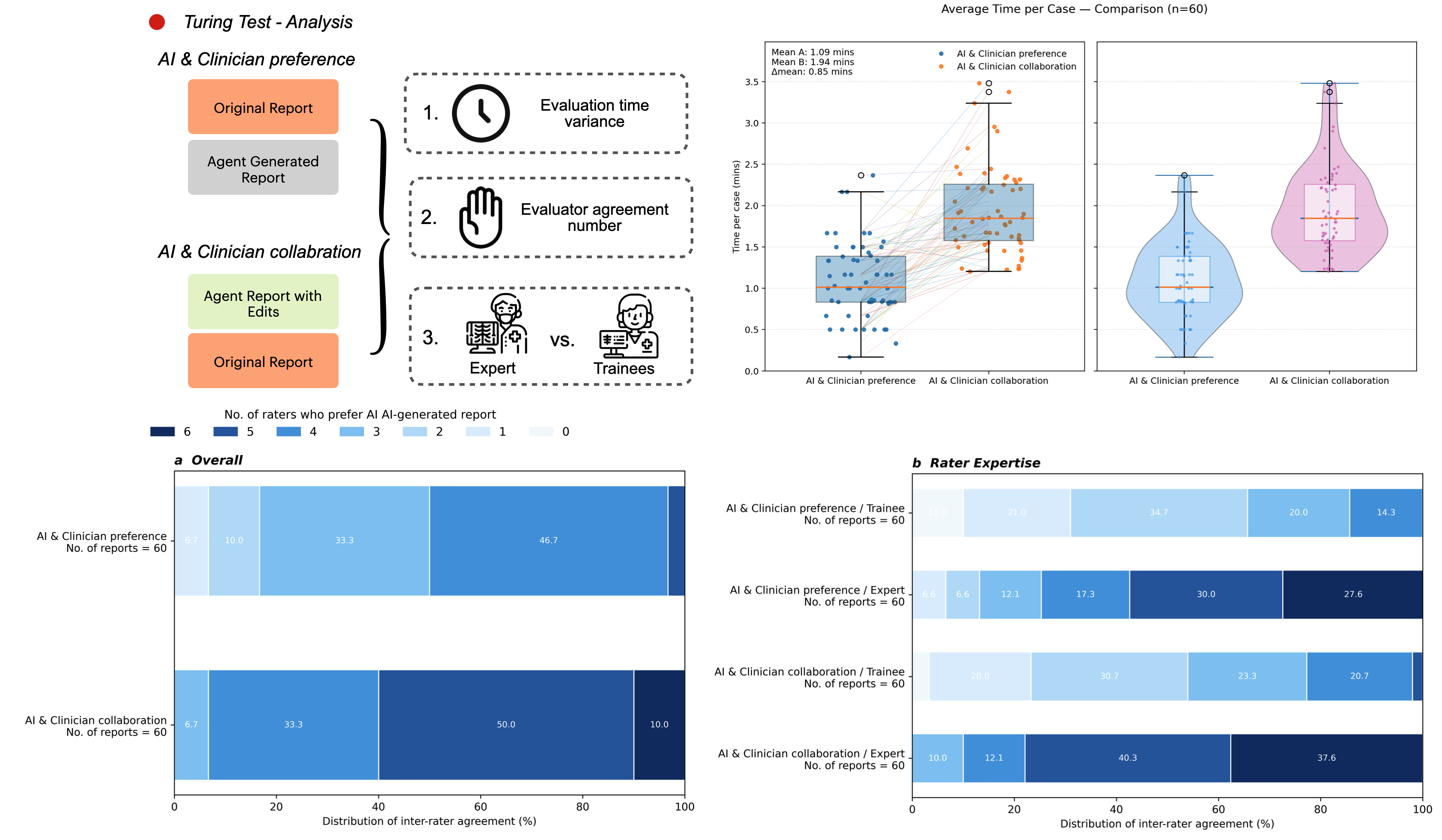}
\caption{\textbf{Analysis of Turing test performance and evaluator agreement.}
\textbf{a} | \textit{Experimental design.} AI-generated radiology reports were evaluated under two settings: (1) \textit{AI \& Clinician preference}, in which radiologists selected between the original and AI-generated reports, and (2) \textit{AI \& Clinician collaboration}, in which a radiologist first edited the AI-generated reports before comparison. Three aspects were analyzed: evaluation time variance, inter-rater agreement, and the effect of rater expertise (expert vs trainee).
\textbf{b} | \textit{Evaluation time per case.} Each point represents one report evaluation ($n=60$). The AI \& Clinician collaboration setting required longer average review time (mean difference $\Delta$ = 0.85 min), reflecting increased depth of review and proofreading effort.
\textbf{c} | \textit{Inter-rater agreement.} Heatmaps display the distribution of reports by the number of raters (0–6) who preferred the AI-generated report, illustrating consistency across evaluators. 
\textbf{d} | \textit{Effect of rater expertise.} Agreement patterns stratified by experience show that experts achieved higher consensus and stronger alignment when reviewing AI-assisted (edited) reports compared with unedited versions. Collectively, these findings indicate that clinician–AI collaboration enhances interpretive consistency and diagnostic trust, albeit with a modest increase in evaluation time.}
\label{turing_test_2}
\end{figure}

\subsection*{Supplementary Methods S1: Cine Loop Standardization Details}
\subsubsection*{1. Rationale and Data Structure}
The input data for our study consists exclusively of ultrasound \textbf{cine loops}, which are video sequences capturing a single, continuous anatomical scan (e.g., a sweep of a specific organ or a focused view of a cardiac valve). These videos are variable in length, with the number of frames, $N$, per video differing significantly across acquisitions. Our deep learning model architecture requires a fixed-size input tensor, which we defined as a sequence of 32 frames. Therefore, a robust and deterministic method was required to convert each variable-length cine loop into a standardized 32-frame sequence. The goal was to achieve this standardization while preserving the crucial temporal information contained within the original scan.
\subsubsection*{2. Standardization Algorithm}
Our algorithm addresses two distinct scenarios based on the original length ($N$) of the cine loop.
\begin{itemize}
    \item \textbf{Case 1: Loops with 32 or more frames ($N \ge 32$)}

    For a cine loop containing $N$ frames, we employ a \textbf{uniform temporal sampling} strategy. This method ensures that the entire temporal span of the acquisition is represented in the final sequence. We select 32 frames at evenly spaced indices from the original video. The index $i_k$ of the $k$-th frame to be selected (where $k$ ranges from 0 to 31) is calculated as:
    
    $$ i_k = \left\lfloor k \times \frac{N-1}{31} \right\rfloor $$

    This deterministic formula guarantees that the first frame ($k=0, i_0=0$) and the last frame ($k=31, i_{31}=N-1$) are always included, with the remaining 30 frames sampled from intervals distributed evenly between them. This approach avoids random selection and ensures that both early and late scanning phases are consistently captured.

    \item \textbf{Case 2: Loops with fewer than 32 frames ($N < 32$)}

    For the minority of cine loops that are shorter than 32 frames, down-sampling is not possible. To meet the 32-frame input requirement without discarding these valuable data points, we perform \textbf{cyclical temporal padding}.

    \begin{enumerate}
        \item First, we take all $N$ original frames in their recorded order.
        \item Then, we append frames to the end of this sequence by restarting from the beginning of the clip (frame 1, frame 2, etc.) until the total length reaches 32.
    \end{enumerate}
    
    For example, a 20-frame sequence \texttt{[f\_1, f\_2, ..., f\_20]} would be padded to \texttt{[f\_1, ..., f\_20, f\_1, ..., f\_12]}. This method maintains the standardized input length while preserving the original temporal coherence of the short clip as much as possible.

\end{itemize}
This comprehensive standardization pipeline ensures that every cine loop, regardless of its original length, is transformed into a 32-frame sequence that serves as a consistent and temporally representative input for our model.

\subsection*{Supplementary Methods S2: Detailed Ablation Study of the Abnormality Classifier}

To dissect the source of the performance gains reported in the main text, we conducted a staged ablation analysis, which revealed a strictly monotonic improvement from the ResNet-50 baseline to the full model (Supplementary Table S2). The substitution of the generic backbone with the domain-pretrained USFM contributed the largest individual increase in macro AUROC (+8.6\%), emphasizing the central role of large-scale, ultrasound-specific representation learning in this task. Each subsequent component—including frame-level attention pooling to generate stable patient-level embeddings, uncertainty-weighted multi-task optimization to balance disease-specific contributions, gradient surgery to resolve conflicting supervision signals, and dynamically learned classification thresholds to adjust for phenotype prevalence and visibility—contributed additive, non-redundant improvements. Altogether, these innovations culminated in a +14.2\% gain in macro AUROC and a +32.7\% gain in macro F1 relative to the baseline.

\begin{table*}[htbp]
\centering
\scriptsize
\caption{\textbf{Comparison of ResNet50 and USFM across multiple diseases.} 
Metrics include AUROC, Average Precision (AP), and F1-score with their differences ($\Delta$).}
\label{tab:resnet_usfm}
\resizebox{\textwidth}{!}{%
\begin{tabular}{l c c c c c c c c c}
\toprule
\textbf{Disease} & 
\multicolumn{3}{c}{\textbf{AUROC}} & 
\multicolumn{3}{c}{\textbf{AP}} & 
\multicolumn{3}{c}{\textbf{F1}} \\
\cmidrule(lr){2-4} \cmidrule(lr){5-7} \cmidrule(lr){8-10}
& ResNet50 & USFM & $\Delta$ 
& ResNet50 & USFM & $\Delta$ 
& ResNet50 & USFM & $\Delta$ \\
\midrule
Cholelithiasis & 0.706 & 0.898 & +0.192 & 0.752 & 0.928 & +0.177 & 0.706 & 0.836 & +0.131 \\
Gallbladder Wall Thickening & 0.672 & 0.854 & +0.182 & 0.331 & 0.610 & +0.280 & 0.418 & 0.597 & +0.179 \\
Renal Cyst & 0.719 & 0.885 & +0.166 & 0.312 & 0.626 & +0.314 & 0.331 & 0.556 & +0.225 \\
Biliary Sludge & 0.728 & 0.818 & +0.090 & 0.522 & 0.724 & +0.202 & 0.568 & 0.657 & +0.090 \\
Gallbladder Distention & 0.656 & 0.743 & +0.088 & 0.468 & 0.534 & +0.066 & 0.537 & 0.587 & +0.050 \\
Medical Renal Disease & 0.780 & 0.866 & +0.086 & 0.195 & 0.353 & +0.158 & 0.218 & 0.383 & +0.165 \\
Common Bile Duct Dilation & 0.686 & 0.771 & +0.085 & 0.228 & 0.449 & +0.221 & 0.290 & 0.397 & +0.107 \\
Cholecystitis & 0.696 & 0.778 & +0.082 & 0.230 & 0.332 & +0.102 & 0.255 & 0.291 & +0.035 \\
Increased Renal Echogenicity & 0.772 & 0.846 & +0.073 & 0.234 & 0.409 & +0.175 & 0.246 & 0.421 & +0.175 \\
Right Pleural Effusion & 0.846 & 0.917 & +0.071 & 0.371 & 0.675 & +0.303 & 0.417 & 0.581 & +0.163 \\
Pericholecystic Fluid & 0.718 & 0.763 & +0.045 & 0.247 & 0.281 & +0.034 & 0.319 & 0.360 & +0.040 \\
Hepatic Steatosis & 0.892 & 0.918 & +0.027 & 0.781 & 0.825 & +0.043 & 0.687 & 0.749 & +0.063 \\
Increased Hepatic Echogenicity & 0.860 & 0.875 & +0.015 & 0.735 & 0.738 & +0.003 & 0.676 & 0.715 & +0.039 \\
Hepatomegaly & 0.735 & 0.749 & +0.014 & 0.467 & 0.518 & +0.051 & 0.496 & 0.520 & +0.024 \\
Cirrhosis & 0.868 & 0.862 & -0.006 & 0.436 & 0.401 & -0.035 & 0.390 & 0.347 & -0.043 \\
Ascites & 0.897 & 0.890 & -0.006 & 0.625 & 0.677 & +0.052 & 0.501 & 0.586 & +0.085 \\
Coarse Hepatic Echotexture & 0.733 & 0.714 & -0.019 & 0.302 & 0.170 & -0.132 & 0.319 & 0.249 & -0.070 \\
Pancreas Poorly Visualized & 0.731 & 0.684 & -0.048 & 0.543 & 0.546 & +0.003 & 0.596 & 0.561 & -0.035 \\
\midrule
\textbf{Macro Average} & 0.761 & 0.824 & +0.063 & 0.432 & 0.544 & +0.112 & 0.443 & 0.522 & +0.079 \\
\bottomrule
\end{tabular}%
}
\end{table*}

\begin{table}[htbp]
\centering
\caption{\textbf{Cross-Site Performance of the Disease Classification Model.} Comparison of Area Under the Receiver Operating Characteristic curve (AUROC), Average Precision (AP), and disease prevalence (Prev.) across the internal Johns Hopkins (JHU) dataset and two external validation cohorts (Stanford and Chinese).}
\label{tab:cross_site_performance}
\resizebox{\textwidth}{!}{%
\begin{tabular}{@{}lccccccccc@{}}
\toprule
\textbf{Disease} & \multicolumn{3}{c}{\textbf{Internal (JHU)}} & \multicolumn{3}{c}{\textbf{Stanford}} & \multicolumn{3}{c}{\textbf{Chinese}} \\ 
\cmidrule(lr){2-4} \cmidrule(lr){5-7} \cmidrule(l){8-10}
 & \textbf{AUROC} & \textbf{AP} & \textbf{Prev.} & \textbf{AUROC} & \textbf{AP} & \textbf{Prev.} & \textbf{AUROC} & \textbf{AP} & \textbf{Prev.} \\ \midrule
Cholelithiasis & 0.895 & 0.917 & 55.5\% & 0.886 & 0.885 & 41.7\% & 0.689 & 0.978 & 95.3\% \\
Pancreas poorly visualized & 0.670 & 0.454 & 32.3\% & 0.694 & 0.373 & 21.3\% & 0.625 & 0.082 & 5.1\% \\
Biliary sludge & 0.818 & 0.733 & 30.5\% & 0.737 & 0.685 & 26.9\% & 0.653 & 0.164 & 7.4\% \\
Gallbladder distention & 0.738 & 0.558 & 32.1\% & 0.670 & 0.345 & 15.7\% & 0.694 & 0.237 & 5.8\% \\
Hepatomegaly & 0.772 & 0.520 & 24.1\% & 0.754 & 0.708 & 42.6\% & 0.566 & 0.095 & 7.0\% \\
Increased hepatic echogenicity & 0.871 & 0.696 & 21.8\% & 0.731 & 0.498 & 25.9\% & 0.689 & 0.568 & 35.8\% \\
Hepatic steatosis & 0.904 & 0.765 & 21.5\% & 0.781 & 0.803 & 43.5\% & 0.728 & 0.683 & 42.0\% \\
Gallbladder wall thickening & 0.823 & 0.583 & 19.1\% & 0.795 & 0.490 & 22.2\% & 0.555 & 0.626 & 53.3\% \\
Ascites & 0.904 & 0.677 & 11.3\% & 0.977 & 0.769 & 7.4\% & 0.794 & 0.214 & 2.7\% \\
Renal cyst & 0.826 & 0.574 & 11.8\% & 0.827 & 0.425 & 8.3\% & 0.681 & 0.415 & 16.3\% \\
Right pleural effusion & 0.926 & 0.712 & 11.1\% & 0.923 & 0.681 & 8.3\% & 1.000 & 1.000 & 0.4\% \\
Common bile duct dilation & 0.775 & 0.327 & 11.0\% & 0.675 & 0.326 & 12.0\% & 0.558 & 0.210 & 16.3\% \\
Pericholecystic fluid & 0.782 & 0.313 & 10.1\% & 0.749 & 0.293 & 8.3\% & 0.792 & 0.027 & 0.8\% \\
Cholecystitis & 0.826 & 0.377 & 8.7\% & 0.548 & 0.075 & 4.6\% & 0.570 & 0.998 & 100.0\% \\
Coarse hepatic echotexture & 0.745 & 0.248 & 8.6\% & 0.713 & 0.156 & 7.4\% & 0.508 & 0.075 & 5.8\% \\
Increased renal echogenicity & 0.845 & 0.503 & 7.6\% & 0.933 & 0.683 & 3.7\% & 0.574 & 0.114 & 7.0\% \\
Cirrhosis & 0.895 & 0.518 & 7.2\% & 0.775 & 0.361 & 7.4\% & 0.629 & 0.047 & 1.6\% \\
Medical renal disease & 0.860 & 0.458 & 5.8\% & 0.886 & 0.692 & 2.8\% & 0.615 & 0.090 & 4.3\% \\ \bottomrule
\end{tabular}%
}
\end{table}

\begin{table}[!htbp]
\centering
\caption{Report generation performance of all model variants across internal (JHU) and external validation datasets (Stanford and Chinese). }
\label{tab:report_generation}
\resizebox{\textwidth}{!}{
\begin{tabular}{lcccccc}
\toprule
\textbf{Model (Dataset)} & \textbf{BLEU-1} & \textbf{BLEU-4} & \textbf{ROUGE-1} & \textbf{ROUGE-L} & \textbf{METEOR} & \textbf{BERT-F1} \\
\midrule
LLaMA3 (JHU)        & 0.5150 & 0.2662 & 0.5535 & 0.4154 & 0.4609 & 0.9395 \\
Phi3 (JHU)          & 0.5210 & 0.2748 & 0.5695 & 0.4369 & 0.4594 & 0.9376 \\
Qwen2.5-VL (JHU)    & \textbf{0.5902} & \textbf{0.3940} & \textbf{0.6572} & \textbf{0.5527} & \textbf{0.5506} & \textbf{0.9494} \\
\midrule
LLaMA3 (Stanford)   & 0.4124 & 0.1339 & 0.5126 & 0.3533 & 0.3866 & 0.8786 \\
Phi3 (Stanford)     & 0.4112 & 0.1464 & 0.5108 & 0.3571 & 0.3878 & 0.8816 \\
Qwen2.5-VL (Stanford) & \textbf{0.4770} & \textbf{0.1546} & \textbf{0.5317} & \textbf{0.4100} & \textbf{0.3625} & \textbf{0.8953} \\
\midrule
LLaMA3 (Chinese)    & 0.3440 & 0.0792 & 0.4475 & 0.2939 & 0.2989 & 0.8628 \\
Phi3 (Chinese)      & 0.4214 & 0.1309 & 0.4906 & 0.3231 & 0.3248 & 0.9114 \\
Qwen2.5-VL (Chinese) & \textbf{0.5040} & \textbf{0.2738} & \textbf{0.6061} & \textbf{0.4286} & \textbf{0.4217} & \textbf{0.9384} \\
\bottomrule
\end{tabular}
}
\end{table}

\begin{table*}[!htbp]
\centering
\caption{Clinically-Focused Evaluation of Report Generation Models.}
\label{tab:clinical_report_gen}
\resizebox{0.85\textwidth}{!}{%
\begin{tabular}{l|c|ccccc}
\toprule
\textbf{Model (Dataset)} & \textbf{DocLens} & \textbf{Degree} & \textbf{Landmark} & \textbf{Feature} & \textbf{Impression} & \textbf{FORTE Avg.} \\
\midrule
\multicolumn{7}{l}{\textit{\textbf{Internal Validation (JHU)}}} \\
Llama-based & 0.4502 & 0.5651 & 0.6165 & 0.5277 & 0.5724 & 0.5704 \\
Phi-based & 0.4477 & 0.5509 & 0.5983 & 0.5174 & 0.5539 & 0.5551 \\
\textbf{Qwen-based} & \textbf{0.6094} & \textbf{0.6462} & \textbf{0.7085} & \textbf{0.6302} & \textbf{0.6380} & \textbf{0.6557} \\
\midrule
\multicolumn{7}{l}{\textit{\textbf{External Validation (Chinese)}}} \\
Phi-based & 0.3111 & 0.4978 & 0.4924 & 0.4214 & 0.6255 & 0.5093 \\
Llama-based & 0.3162 & 0.4955 & 0.4835 & 0.4015 & 0.6314 & 0.5030 \\
\textbf{Qwen-based} & \textbf{0.3843} & \textbf{0.5792} & \textbf{0.6093} & \textbf{0.5529} & \textbf{0.7311} & \textbf{0.6181} \\
\midrule
\multicolumn{7}{l}{\textit{\textbf{External Validation (Stanford)}}} \\
Llama-based & 0.3821 & 0.5333 & 0.5459 & 0.4306 & 0.6205 & 0.5326 \\
Phi-based & 0.3908 & \textbf{0.5551} & 0.5531 & \textbf{0.4540} & \textbf{0.5929} & \textbf{0.5388} \\
\textbf{Qwen-based} & \textbf{0.5470} & 0.5444 & \textbf{0.5563} & 0.4481 & 0.5524 & 0.5253 \\
\bottomrule
\end{tabular}%
}
\end{table*}

\begin{table*}[htbp] 
\centering
\scriptsize

\caption{
    \textbf{Prevalence of Medical Findings.}
    Top findings sorted by frequency, with cases, prevalence, and clinical significance.
}
\label{tab:findings_prevalence}
\begin{tabular}{c l c c l}
\toprule
\textbf{Rank} & \textbf{Disease} & \textbf{Cases} & \textbf{Prevalence} & \textbf{Clinical Significance} \\
\midrule
1  & Cholelithiasis & 5,390 & 54.1\% & Most common biliary pathology; risk of biliary colic and cholecystitis. \\
2  & Pancreas poorly visualized & 3,410 & 34.3\% & Technical limitation that reduces diagnostic sensitivity for pancreatic disease. \\
3  & Biliary sludge & 3,167 & 31.8\% & Precursor to gallstones; may cause biliary colic or pancreatitis. \\
4  & Gallbladder distention & 3,011 & 30.2\% & Can reflect fasting state, obstruction, or dyskinesia. \\
5  & Hepatomegaly & 2,489 & 25.0\% & Suggests hepatic disease, congestion, or infiltration. \\
6  & Increased hepatic echogenicity & 2,168 & 21.8\% & Often correlates with steatosis and possible fibrosis. \\
7  & Hepatic steatosis (fatty liver) & 2,162 & 21.7\% & Common metabolic liver disease; cardiometabolic risk marker. \\
8  & Gallbladder wall thickening & 2,094 & 21.0\% & Supports inflammatory processes (e.g., cholecystitis) or systemic edema. \\
9  & Ascites & 1,329 & 13.4\% & Associated with portal hypertension, malignancy, or infection. \\
10 & Common bile duct dilation & 1,140 & 11.5\% & Suggests biliary obstruction (stone, stricture, or mass). \\
11 & Renal cyst & 1,086 & 10.9\% & Usually benign; further evaluation if complex features present. \\
12 & Pericholecystic fluid & 1,044 & 10.5\% & Ancillary sign supporting acute cholecystitis. \\
13 & Right pleural effusion & 1,005 & 10.1\% & Can occur with heart failure, hepatic hydrothorax, or infection. \\
14 & Cholecystitis & 844 & 8.5\% & Acute inflammation of the gallbladder; requires prompt management. \\
15 & Coarse hepatic echotexture & 775 & 7.8\% & Imaging correlate of chronic liver disease/fibrosis. \\
16 & Increased renal echogenicity & 696 & 7.0\% & Suggestive of medical renal (parenchymal) disease. \\
17 & Cirrhosis & 660 & 6.6\% & Advanced chronic liver disease with portal hypertension risk. \\
18 & Medical renal disease & 510 & 5.1\% & Non-obstructive parenchymal kidney disease. \\
19 & Positive Murphy's sign & 497 & 5.0\% & Sonographic tenderness supporting acute cholecystitis. \\
20 & Hepatic lesion/mass & 477 & 4.8\% & Requires characterization (benign vs malignant). \\
21 & Intrahepatic bile duct dilation & 459 & 4.6\% & Indicates intrahepatic obstruction or cholestasis. \\
22 & Hepatic cyst & 389 & 3.9\% & Typically simple and benign; follow if atypical. \\
23 & Gallbladder polyp & 364 & 3.7\% & Malignancy risk increases with size; surveillance per guidelines. \\
24 & Gallbladder adenomyomatosis & 313 & 3.1\% & Benign hyperplastic change; characteristic comet-tail artifacts. \\
25 & Gallbladder wall hyperemia & 274 & 2.8\% & Doppler sign supporting acute cholecystitis. \\
26 & Renal atrophy & 255 & 2.6\% & Suggests chronic kidney disease or prior ischemic insult. \\
27 & Nephrolithiasis & 254 & 2.6\% & Renal stones; risk of obstruction and infection. \\
28 & Hepatic hemangioma & 222 & 2.2\% & Common benign vascular liver tumor. \\
29 & Pancreatic duct dilation & 190 & 1.9\% & May reflect chronic pancreatitis or obstructing lesion. \\
30 & Hydronephrosis & 181 & 1.8\% & Sonographic sign of obstructive uropathy. \\
31 & Choledocholithiasis & 144 & 1.4\% & Stones in the common bile duct; cholangitis/jaundice risk. \\
32 & Pancreatitis & 130 & 1.3\% & Inflammation of the pancreas; correlate clinically/labs. \\
33 & Renal cortical thinning & 124 & 1.2\% & Chronic parenchymal loss; CKD indicator. \\
34 & Perinephric fluid & 113 & 1.1\% & Seen with inflammation, trauma, or urinoma. \\
35 & Portal hypertension & 108 & 1.1\% & Elevated portal pressure; varices and ascites risk. \\
36 & Calcified granuloma & 101 & 1.0\% & Healed prior infection; typically incidental. \\
37 & Decreased portal vein velocity & 77  & 0.8\% & Hemodynamic sign suggestive of portal hypertension. \\
38 & Hepatofugal flow & 61  & 0.6\% & Reversed portal flow; advanced portal hypertension. \\
39 & Pancreati   c cyst & 60  & 0.6\% & Spectrum from benign to neoplastic; follow-up per size/features. \\
40 & Suspicious renal mass & 48  & 0.5\% & Possible malignancy; cross-sectional imaging/urology referral. \\
41 & Portal vein occlusion & 23  & 0.2\% & Thrombosis/occlusion; portal hypertension and ischemia risk. \\
42 & IVC obstruction & 5   & 0.1\% & Impaired venous return; evaluate for thrombus/compression. \\
\bottomrule
\end{tabular}
\end{table*}

\begin{table*}[t]
\centering
\caption{RUQ Ultrasound FORTE keyword list grouped by category. Synonyms appear as slash-separated forms.}
\label{tab:ruq_forte_keywords}
{\fontsize{8}{9.6}\selectfont
\renewcommand{\arraystretch}{1.05}
\setlength{\tabcolsep}{4pt}
\resizebox{\textwidth}{!}{%
\begin{tabular}{@{}l@{}}

\begin{tabular}{@{}p{0.24\textwidth} p{0.24\textwidth} p{0.24\textwidth} p{0.24\textwidth}@{}}
\midrule
\multicolumn{4}{l}{\textbf{Degree}}\\
\midrule
abnormality/ sign & amount & apparent & common \\
completely & consistent & atrophic/ contracted/ decreased/ decompressed/ nondilated/ underdistended & diffuse/ diffusely \\
distally & early & end-stage & dilation/ distended/ distension/ enlarged/ prominent \\
discrete/ focal/ focally/ foci/ focus & homogeneous & hypoattenuation & increased/ large/ largest/ maximal/ maximum/ peak/ top/ upper \\
inferior & limited/ poorly/ suboptimally & local/ localized/ territorial & mild/ minimally/ mildly/ slightly \\
moderate & multiple/ numerous & absent/ negative/ no/ not & healthy/ intact/ normal/ preserved/ unremarkable \\
pathologically & indeterminate/ possibly/ suggestion & present & relatively \\
minimal/ slightly/ small/ smaller/ trace & subcentimeter & thickened/ thickness &  \\
\end{tabular}

\\[2pt]\midrule

\begin{tabular}{@{}p{0.24\textwidth} p{0.24\textwidth} p{0.24\textwidth} p{0.24\textwidth}@{}}
\multicolumn{4}{l}{\textbf{Landmark}}\\
\midrule
anatomic & antegrade/ hepatopedal & arterial/ aorta/ aortic/ hyperemic/ vascular & bile duct/ biliary \\
calyces & cell & contour & corticomedullary/ cortical \\
cystic & decubitus & erect & evidence/ present \\
exophytic & extrahepatic & flank & fossa \\
fundus & gallbladder & grayscale & habitus \\
head & internal/ intraluminal & intervening & intrahepatic \\
inferior vena cava/ IVC/ venous/ vein & kidney/ renal/ urinary & left & AP/ craniocaudal/ dimension/ length/ longitudinal \\
hepatic/ hepatocellular/ liver & bowel/ loops & duct/ lumen/ tract & lymph \\
main & interpolar/ mid/ midportion & midclavicular &  \\
neck & pancreas/ pancreatic & body/ cortical/ parenchyma/ parenchymal & pathologically \\
adjacent/ pericholecystic/ perinephric & peripancreatic & pleural & pole \\
polyp & portal vein & portion & posterior \\
postprandial & quadrant & report &  \\
reversed & Riedel's lobe & right & sagittal \\
septation/ volume & obscuration/ obscured/ ringdown/ shadow/ shadowing & size & sludge \\
solid & splenule & stent & stranding \\
structure & superomedial & surrounding & tail \\
tapering & terminate & Doppler/ sonographic/ sonographically/ transducer/ ultrasound & tree \\
umbilical & underlying & undulating & variant \\
velocity & wall &  &  \\
\end{tabular}

\\[2pt]\midrule

\begin{tabular}{@{}p{0.24\textwidth} p{0.24\textwidth} p{0.24\textwidth} p{0.24\textwidth}@{}}
\multicolumn{4}{l}{\textbf{Feature}}\\
\midrule
acoustic & anechoic & aneurysm & artifact \\
atherosclerotic & avascular & caliber/ diameter & coarse/ coarsened/ heterogenous \\
color & debris & defect & detected/ demonstrated/ displays/ displayed/ reveals \\
echo/ echotexture & echogenic/ echogenicities/ hyperechoic/ steatosis & elicited & enhancement \\
examination/ interrogation & extent & ascites/ cyst/ edema/ effusion/ fluid & gas/ gaseous/ pneumatosis/ pneumobilia \\
dependent/ layering/ layers & lesion/ mass/ masses & material & measuring \\
medication & mobile & node & nodular \\
nonmobile & nonobstructing & nonshadowing & overload \\
overlying/ superimposing & palpation & nonocclusive/ patent/ recanalized & polyp \\
reactive & ringdown & sand-like/ sandpaperlike & septation \\
obscuration/ obscured/ shadow/ shadowing & size/ volume & sludge & solid \\
stent & calculi/ calcification/ stone & stranding & tapering \\
terminate & Doppler/ sonographic/ sonographically/ ultrasound & underlying & undulating \\
evidence/ visualized & waveform & window &  \\
\end{tabular}

\\[2pt]\midrule

\begin{tabular}{@{}p{0.24\textwidth} p{0.24\textwidth} p{0.24\textwidth} p{0.24\textwidth}@{}}
\multicolumn{4}{l}{\textbf{Impression}}\\
\midrule
adenomyomatosis & carcinoma & cholecystitis & cholelithiasis \\
cirrhosis & disease & hemangioma & hepatomegaly \\
hydronephrosis & inflammation & intussusception & metastasis/ metastatic \\
nephrolithiasis & steatosis & Murphy/ pain/ tender/ tenderness & thrombus/ thrombosis \\
\midrule
\end{tabular}

\end{tabular}%
} 
} 
\end{table*}

\begin{table*}[h!]
\centering
\caption{Summary of evaluator experience for attending radiologist. Evaluator names are anonymized by initials.}
\begin{tabular}{lccc}
\hline
\textbf{Evaluator (Initials)} & \textbf{Professional Title / Role} & \textbf{Years of Experience} & \textbf{Specialty / Expertise} \\
\hline
H. B. & Attending Radiologist & 7 & Diagnostic Radiology \\
F. S. & Attending Radiologist & 8 & Diagnostic Radiology \\
S. K. & Attending Radiologist & 8 & Diagnostic Radiology \\
C. T. & Attending Radiologist & 4 & Diagnostic Radiology \\
C. L. & Attending Radiologist & 12 & Diagnostic Radiology \\
D. O. & Attending Radiologist & 17 & Diagnostic Radiology \\
U. H. & Attending Radiologist & 43 & Diagnostic Radiology \\
J. A. & Attending Radiologist & 8 & Diagnostic Radiology \\
\hline
\end{tabular}
\label{tab:evaluator_experience}
\end{table*}

\begin{table*}[h!]
\centering
\caption{Summary of evaluator experience for radiology resident or fellow. Evaluator names are anonymized by initials.}
\begin{tabular}{lccc}
\hline
\textbf{Evaluator (Initials)} & \textbf{Professional Title / Role} & \textbf{Years of Experience} & \textbf{Specialty / Expertise} \\
\hline
J. S. & Radiology Resident & PGY-2 & Diagnostic Radiology \\
J. F. & Radiology Resident & PGY-3 & Diagnostic Radiology \\
R. W. & Radiology Resident & PGY-4 & Diagnostic Radiology \\
L. Z. & Radiology Fellow & PGY-6 & Diagnostic Radiology \\
\hline
\end{tabular}
\label{tab:evaluator_experience}
\end{table*}

\newpage
\subsection*{Details on ICD Code Selection}

\textbf{Tokyo Guidelines–based Criteria for Acute Cholecystitis (Ultrasound):}
To identify ultrasound examinations related to acute cholecystitis (AC), we selected ICD-10 codes aligned with the Tokyo Guidelines 2013/2018 (TG13/TG18) diagnostic framework \cite{hirota2007diagnostic}.
According to these guidelines, diagnosis requires evidence across three domains—Local Signs of Inflammation, Systemic Signs of Inflammation, and Imaging Findings.
ICD codes corresponding to these categories were used to identify relevant clinical encounters, restricted to those recorded within 15 days before or after the ultrasound examination date to ensure temporal relevance.

\begin{itemize}[itemsep=0pt, topsep=0pt]
\item Murphy’s sign – R10.819 (Right upper quadrant abdominal tenderness)
\item Right upper quadrant mass, pain, or tenderness – R10.811, R10.819, R19.00
\item Fever – R50.9 (Fever, unspecified)
\item Elevated C-reactive protein (CRP) – R79.82
\item Elevated white blood cell count (WBC) – D72.829
\item Acute cholecystitis with gallstones – K81.0
\item Acute cholecystitis without gallstones – K81.9
\item Pericholecystic fluid (associated with gallbladder inflammation) – covered under K81.0
\item Gallstones (cholelithiasis) without cholecystitis – K80.20
\end{itemize}

\end{document}